# Inductance of Circuit Structures for MIT LL Superconductor Electronics Fabrication Process with 8 Niobium Layers

Sergey K. Tolpygo, Vladimir Bolkhovsky, T.J. Weir, C.J. Galbraith, Leonard M. Johnson, *Senior Member, IEEE,* Mark A. Gouker, *Senior Member, IEEE,* and Vasili K. Semenov

*Abstract*—Inductance of superconducting thin-film inductors and structures with linewidth down to 250 nm has been experimentally evaluated. The inductors include various striplines and microstrips, their 90º bends and meanders, interlayer vias, *etc.*, typically used in superconducting digital circuits. The circuits have been fabricated by a fully planarized process with 8 niobium layers, developed at MIT Lincoln Laboratory for very-large-scale superconducting integrated circuits. Excellent run-to-run reproducibility and inductance uniformity of better than 1% across 200-mm wafers have been found. It has been found that the inductance per unit length of stripline and microstrip line inductors continues to grow as the inductor linewidth is reduced deep into the submicron range to the widths comparable to the film thickness and magnetic field penetration depth. It is shown that the linewidth reduction does not lead to widening of the parameter spread due to diminishing sensitivity of the inductance to the linewidth and dielectric thickness. The experimental results were compared with numeric inductance extraction using commercial software and freeware, and a good agreement was found for 3-D inductance extractors. Methods of further miniaturization of circuit inductors for achieving circuit densities > $10^6$ Josephson junctions per $cm^2$ are discussed.

## I. INTRODUCTION

INDUCTORS are the second – after Josephson junctions - most important component of all superconducting digital integrated circuits based on RSFQ [1]-[3], RQL [4]-[5], ERSFQ [6], and other approaches, providing a network for storing and routing flux quanta and single flux quantum (SFQ) pulses generated by Josephson junctions as well as distributing all dc and rf bias and clock signals. Inductors contribute the main fraction to circuit area and largely determine its density. Progress of superconductor electronics and potential applications for energy efficient and high performance computing demand development of superconducting very large scale integrated (VLSI) circuits. This in turn demands reducing linewidth of all circuit components (first of all inductors and resistors) into deep submicron range. Currently, the typical inductor linewidth, $w$, in SFQ circuits is in the range from 1 µm to 5 µm. It is set by the minimum linewidth of fabrication processes available [7]–[10] and circuit design-related limits to the parameter spreads (*e.g.*, inductance standard deviation, $\sigma_L$) on chip, on wafer, and wafer-to-wafer variation.

Recently, we have developed a fully-planarized Nb-based fabrication process for circuits with 8 superconducting layers and Nb/Al-AlO$_x$/Nb Josephson junctions (JJs) with 10 kA/cm$^2$ (100 µA/µm$^2$) critical current density [11]-[13]. This process has yielded circuits with over $7 \cdot 10^4$ JJs on a 5 mm x 5 mm chip. Its 10-metal-layer extension is aimed at fabricating integrated circuits with >$10^6$ JJ/cm$^2$. The current process is based on a 248-nm photolithography, which sets the practical limit on the minimum linewidth and pitch at about 200 nm and 400 nm, respectively, and allows for almost 5x reduction of inductor linewidth in SFQ circuits. If sheet inductance (inductance per square) of a thin-film inductor were independent of its linewidth, the inductor area would scale as $w^2$. Hence, a 5x reduction in $w$ offered by this process would translate into a very significant reduction in the inductor area.

There have been a number of papers devoted to extracting inductances of superconducting structures and a number of different analytical and numeric methods developed for inductance modeling [14]–[30]. To the best of our knowledge, there are no available experimental data on superconducting inductors with linewidths less than about 0.5 µm in SFQ circuits as well as no verification of the existing numerical methods and inductance modeling software for inductors with widths comparable to the penetration depth and the thickness of the films and interlayer dielectric.

In order to fill this gap and provide input parameters for SFQ circuit designers designing into our 8-metal-layer and future 10-metal-layer fabrication processes, we experimentally evaluated inductances of thin-film inductors with linewidth in the range from 250 nm to 4 µm, of various shapes and for the typical combinations of layers available in the process, as well as "parasitic" inductance associated with vias between the superconducting layers. We also provide a comparison of the obtained data with numeric simulations using several software packages [16], [17],[18],[30],[31].

## II. FABRICATION PROCESS, INDUCTOR DESIGN, AND TESTING

### A. 8-metal layer fabrication process

A cross section of our 8-metal-layer (8M) fabrication process is shown in Fig. 1. We used 248-nm photolithography and dry etching in a 200-mm-wafer high density plasma etcher to pattern all metal and dielectric layers. All dielectric layers were planarized using chemical mechanical polishing (CMP). This provided for a flat topography and allowed for the focus depth required for fine patterning of metal layers. The only layer that was not planarized was a 60-nm dielectric layer I5R covering the layer of resistors, R5. Its planarization was not needed because the maximum topography variation on this layer was less than the 40 nm created by the patterned resistors. $SiO_2$ interlayer dielectric was deposited using plasma-enhanced chemical vapor deposition (PECVD).

TABLE I
DESIGN AND FABRICATION PARAMETERS FOR INDUCTORS STUDIED

| Notation | Inductor type | Dielectric thickness, $d$ (nm) | Thickness of metal layers, $t$ (nm) |
|---|---|---|---|
| M6_M7 | Inverted microstrip | 200 | 200_200 |
| M5_M4 | Microstrip | 200 | 135_200 |
| M5_M6 | Inverted microstrip | 280 | 135_200 |
| M6_M5 | Microstrip | 280 | 200_135 |
| M6_M4 | Microstrip | 615 | 200_200 |
| M5_M7 | Inverted microstrip | 680 | 135_200 |
| M4_M6_M7 | Stripline | 615_200 | 200_200_200 |
| M4_M5_M7 | Stripline | 200_680 | 200_135_200 |
| M5_M6_M7 | Stripline | 280_200 | 135_200_200 |
| M4_M6_M7 | 90-degree bend | 615_200 | 200_200_200 |
| M4_M6_M7 | Meander, $s$ = 0.35, 0.5, and 0.7 µm | 615_200 | 200_200_200 |

For microstrips, the thickness of metal layers is shown in the format $t_1\_t_2$, where $t_1$ and $t_2$ are thicknesses of the signal layer and the ground plane, respectively. Similarly for striplines, the format is $t_1\_t_2\_t_3$ with $t_2$ corresponding to signal layer. For dielectric thicknesses, the format is $d_{1,2}\_d_{2,3}$ where $d_{i,j}$ indicates the dielectric thickness between metal layers $i$ and $j$. For combinations involving M5 layer, the dielectric thickness includes also 40-nm anodization layer. The nominal thicknesses are shown. The actual thickness is expected to be within ± 20 nm from the nominal.

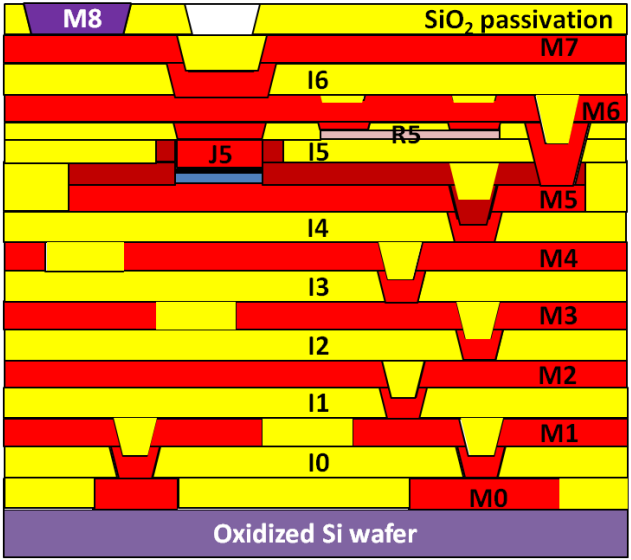

Fig. 1. Cross-section of 8-metal-layer (8M) fabrication process developed at MIT LL and used in this work to fabricate test circuits for extracting inductances of various inductors for SFQ circuits. Layers M0 – M4, M6, and M7 are 200-nm thick Nb layers. M8 is contact pad metallization, Ti/Pt/Au. $SiO_2$ intermetal dielectric layers I0 – I7 are 200-nm thick. J5 is JJ top electrode. M5 is JJ bottom electrode, 135-nm thick, covered by a 40-nm layer of anodization. R5 is 40-nm molybdenum resistor layer. It is covered by a 60-nm layer of $SiO_2$. The total $SiO_2$ thickness between M5 and M6 is 240 nm. The nominal distances between metal layers used for circuit inductors are given in Table 1. The upper four Nb layers, M4 - M7, of this process are used in the truncated 4-metal-layer process.

We used Nb/Al-AlOx/Nb Josephson junctions with 10 kA/cm$^2$ (100 µA/µm$^2$) critical current density and minimum diameter of 500 nm. The junctions' sidewalls were passivated using anodization. The patterned layer of JJs was also fully planarized using dielectric CMP. An SEM image of the fabricated structure cross-sectioned using focused ion beam (FIB) is shown in Fig. 2. More details on the fabrication process are given in [11]-[13].

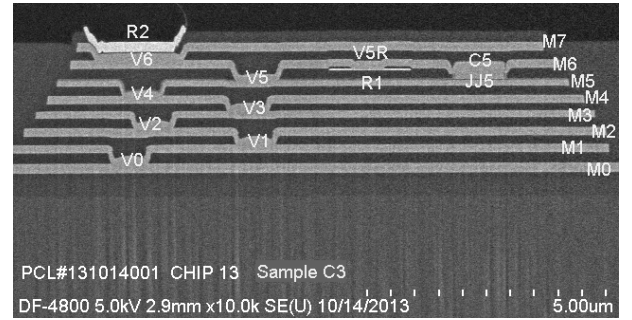

Fig. 2. SEM picture of a cross section of a fully processed 8-metal-layer circuit. All $SiO_2$ dielectric layers were planarized by CMP, resulting in perfectly flat topography of metal layers except in etched vias (marked as V0, V1, *etc.*) between adjacent layers. JJ is marked as JJ5, resistor as R1, contact pad as R2.

For simplicity and statistics, many test structures were also fabricated by a truncated process using only upper layers from M4 through M8. In all other respects this 4-metal-layer (4M) fabrication process and the parameters of physical layers used are identical to the 8M process. So the results obtained on inductance test structures fabricated by the 4M process are fully applicable to the 8M process and vice versa.

### B. Inductor Test Structures

Most often used configurations of inductors in SFQ circuits are microstrips, inverted microstrips, and striplines as shown in Fig. 3(a). They can be bent or meandered to accommodate the available space and required routing, see Fig. 3(c) and Fig. 3(d). Due to a very large number of possible double- and triple-layer combinations and shapes that could be formed using 8 available superconducting layers, we restricted our study to layers in close proximity to the JJ layer, layers M4 to M7, which are of prime importance for SFQ circuit design. Results on the bottom layers M0 to M3 will be reported elsewhere.

Design and fabrication parameters of the studied structures are shown in Table 1. We will refer to microstrip and inverted microstrip line inductors using Signal_Ground notation, e.g., M6_M7 identifying signal layer, M6, and then the ground plane layer, M7. Similarly, all stripline configurations will be referred to as Ground_Signal_Ground, *e.g.*, M4_M6_M7 identifying the patterned signal line with width $w$ on layer M6 sandwiched between two ground planes: M4 below it and M7 above it.

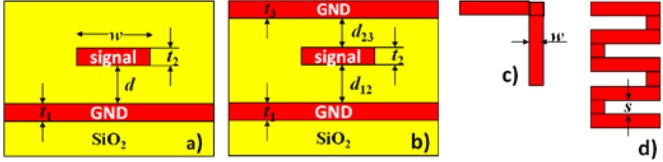

Fig. 3. Cross sections of microstrip (a) and stripline (b) inductors. Top view of the signal line with a 90-degree inductor bend (c) and inductor meander (d). Signal linewidth is $w$ and $s$ is space; meander pitch $p = w + s$. The linewidth was varied from 0.25 µm to 4 µm, and space from 0.35 µm to 0.70 µm.

Connections to signal lines and between ground planes are provided by vias etched in interlayer dielectrics and filled by Nb during metal layer deposition, see Fig. 1 and Fig. 2. These superconducting connections have some inductance, which is often termed 'parasitics,' that needs to be measured and characterized for accurate circuit design. Because of the very large number of possible via configurations, in this work we have characterized only a few of the most typical ones.

*C. Circuit design and testing*

Inductance of superconducting components can be conveniently extracted by including them in a SQUID loop and measuring the period of the SQUID modulation. The measured period is related to the total SQUID inductance, a part of which is the inductor device under test (DUT). Seemingly simple in principle, separation of these parts is often difficult, (*e.g.*, requires measuring multiple devices with varying length of the inductor), and the measurements often suffer from parasitics related to connections between the SQUID and DUT, mutual inductance between the DUT and magnetic bias lines, *etc*.

In order to minimize such effects and also maximize the number of DUTs on a 5 mm x 5 mm chip with a limited number of I/O contact pads, we developed a circuit which allows for differential measurements of multiple inductors and requires 1+ε contact pads for extraction of one inductive parameter. The efficiency is achieved because most of the contact pads are shared between several similar experiments as shown in Fig. 4. The circuit requires $N+2$ contact pads to bias, with respect to the common ground, $N$ SQUIDs and measure their voltage-flux characteristics, so ε = 2/$N$.

We used an array of similar SQUIDs which differ only by value (and or type) of inductors in their arms. The circuit diagram is shown in Fig. 4 along with the layout of one of the SQUIDs.

Bias current iBias is equally split between the SQUIDs using resistors RI. The value of these resistors should be large enough to prevent undesirable cross talk between the SQUIDs.

The nature of the cross talk is trivial - during flux-voltage $V(\Phi)$ measurements SQUIDs are in resistive states and behave as nonlinear resistors affecting the distribution of bias current between the SQUIDs. The cross talk is small if the nonlinear SQUID resistances are much less than RI. The voltage drop on each SQUID (with respect to the common ground) can be measured individually using terminals sqOut1, sqOut2, *etc.*, coming out to I/O contact pads at the chip perimeter.

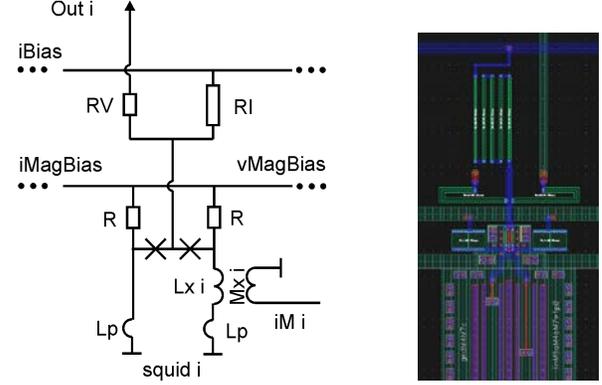

Fig. 4. Schematics of a SQUID array used for extracting inductance of various inductors (left panel) and a layout of one of the SQUIDs (right panel).

Magnetic flux in each SQUID is created by passing electrical currents via corresponding inductors. For example, for SQUIDi (Fig. 4), these inductors are Lxi and Lp. The main trick here is that equal currents flow via the left and the right arms of the SQUID. As a result, the induced flux is proportional to the difference between inductances of the left and right arms of the SQUID. All other (parasitic) inductances associated with connections to the arms, to the ground, the SQUID, *etc.*, are made equal. The left arm layout is a mirror of the right arm layout, except for the difference we want to measure (DUT). As a result, the flux applied to the SQUID is proportional to Lxi.

Since the external iMagBias current is equally split between arms of all the SQUIDs in the array, the flux in one SQUID is

$$\Phi = L_{xi} \cdot \text{iMagBias}/2N, \qquad (1)$$

where $N$ is number of SQUIDs in the array. This allows us to find the difference of the two inductors or Lxi that differ only by the length, $l$, by measuring the period of the SQUID modulation by iMagBias. We used designs with $N = 6$ and $N = 7$ in this work, with total number of DUT of 24 per 5 mm x 5 mm chip.

The described technique allows further generalizations for measuring mutual inductance, inductance of vias, parasitic inductance, *etc.* For instance, the right arm of SQUIDi in Fig. 4 is magnetically coupled to wire Mi. The mutual inductance Mxi can be extracted by scanning current iMi applied to wire Mi and measuring the period of modulation of SQUIDi.

*D. Simulations*

Most of the designed inductors were simulated using the available software such as InductEx [16], Sonnet [31], and [30] in order to extract inductance per unit length, $L/l$, for microstrips and striplines, and inductance per 90-degree bend associated with corner square shown in Fig. 3(c) for a single bend, and the difference between the straight and meander inductors of the same length. In the simulations, the same magnetic field penetration depth $\lambda = 90$ nm was assumed for all superconducting layers used in the process, and the nominal thicknesses of the metal and dielectric layers were used, shown in Table 1. The simulated results were compared with the experimental data. No attempt was made to achieve the best fit to the data either by adjusting the penetration depth or including the possible deviation in the actual thickness.

### III. RESULTS

The typical experimental results for straight microstrip inductors are shown in Fig. 5. Solid lines show results of simulations using InductEx [16]. The difference between the data and the simulations is within 2% at large linewidths and increases to 5% - 7% for the smallest widths. Similarly good agreement was found for a different 3-D inductance extraction package, Program LL developed in [17],[18],[30], see dashed lines in Fig. 5 and in Fig. 6(a).

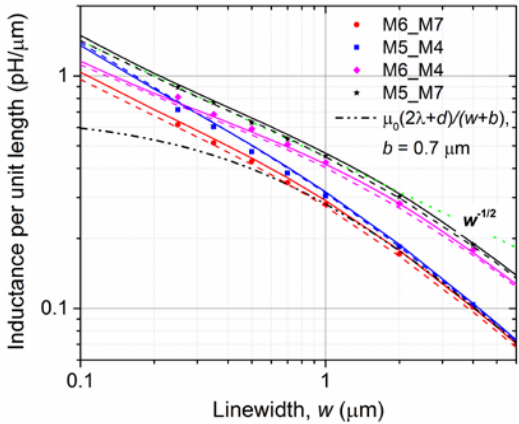

Fig. 5. Microstrip inductance per unit length, $L/l$, for microstrip inductors with length $l \gg w$ and the most typical combinations of circuit layers. The error of the measurements is about the width of the data points. Solid lines show simulations using InductEx with $\lambda = 90$ nm. Dashed lines show simulations using Khapaev's inductor extractor, Program LL [18],[30]. Dash-dotted line is an approximation due to Chang [19]-[21] with a 'fringing' factor $b = 0.7$ µm, fitting the M6_M7 data only at large widths $w > 1$ µm. In the entire range 0.25 µm $\leq w \leq$ 4 µm studied, the maximum difference between the simulated $L/l$ and the experiment is less than 7%, being the largest at the smallest linewidths.

Fig. 7 shows the typical results for stripline inductors. We find a reasonable agreement of 3-D inductance extractors [30] and [16] with the data, although not as good as for the microstrips. 2-D+ simulations, e.g., using Sonnet [31], show a large difference from the data at small $w$. We note that, for simple geometries such as striplines, Program LL [30] gives a better agreement with the experimental data than InductEx, see Fig. 7, which tends to overestimate the inductances. However, InductEx is a much more capable tool for more complex 3-D geometries including bends, meanders, vias, *etc.*, which cannot be handled by [30].

Uniformity of inductances was assessed by measuring the same test structures on 8 or 9 dies across 200-mm wafers, as shown in Fig. 6 (right panel). The typical data for the 8M process run are shown in Table II. Run-to-run variation of the mean inductance is within ± 2% for $w > 0.7$ µm, increasing to ±6% for the narrowest width studied.

TABLE II
INDUCTANCE UNIFORMITY FOR M6_M4 MICROSTRIPS ON 200-MM WAFER

| Width, $w$ (µm) | Mean inductance per unit length, (pH/µm) | Standard deviation, $1\sigma$, (%) |
|---|---|---|
| 0.5 | 0.666(7) | 1.5 |
| 1.0 | 0.442(2) | 1.2 |
| 2.0 | 0.292(6) | 0.7 |
| 4.0 | 0.170(3) | 0.5 |

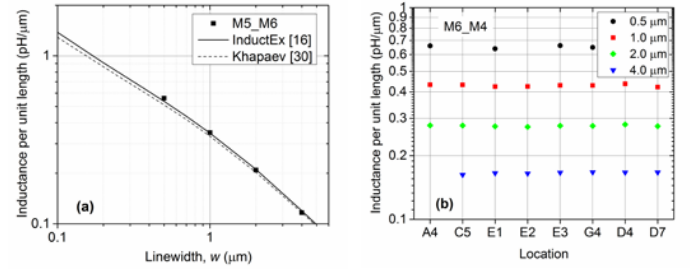

Fig. 6. Inductance of M5_M6 microstrips at different linewidth (a), and inductance uniformity across 200-mm wafers for M6_M4 microstrips (b). Locations on wafer are labeled A4, C5, *etc.*, corresponding to 22 mm x 22 mm stepper exposure fields on a 7x7 grid (A, B,…,G; 1,2…,7).

We find that, in all the cases, inductance per unit length at narrow widths scales as $1/w^{1/2}$ rather than $1/(w+b)$ as is the case at $w \gg d,t,\lambda$ [19]-[21], where $d$ and $t$ are the dielectric gap and signal line thickness, respectively. This is actually to our advantage as shrinking the linewidth allows for increasing the circuit density by decreasing the inductance area roughly as $w^{3/2}$. Additionally, this decreasing sensitivity of the inductance to the inductor linewidth helps in getting the parameter spreads tighter, even for the linewidths approaching the resolution limit of the 248-nm photolithography where ±10% variation of the linewidth is typical.

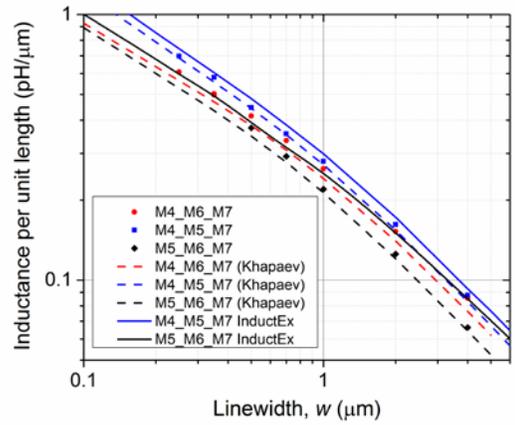

Fig. 7. Stripline inductance per unit length of stripline inductors with $l \gg w$ for several combinations of the signal and ground plane layers. Color-coded dashed and solid lines show simulations using LL [30] and InductEx [16], respectively. The color of the lines corresponds to the color of the data points.

Similarly, as the inductor linewidth decreases, we see a diminishing sensitivity of the inductance to the dielectric thickness. For example, let us compare M5_M7 and M6_M7 microstrips, which differ by a factor of 3.4x in the dielectric thickness. This difference translates into a factor of 1.9x in the inductance at $w = 4$ µm and reduces to 1.4x at $w = 0.25$ µm. In some sense, all inductors at very narrow widths become nearly identical. Indeed, at $w, t << d$ and $w \approx t$, the microstrip becomes just a thin wire above the ground plane. In this case, the geometric inductance depends logarithmically on the distance $L/l = (\mu_0/2\pi)\ln(4d/t)$ and the total inductance is dominated by the kinetic part. This fact helps in keeping the inductance spreads low despite unavoidable variations of local dielectric thickness, which are caused by chemical mechanical polishing (CMP) for planarization. The typical thickness variation we observe in our CMP process is in the range ±20 nm for a 200-nm layer, whereas inductance variation on 200-mm wafers is less than a few percent, see Table II.

## IV. CONCLUSION

We have measured inductance of various superconducting microstrip-line and stripline inductors, their bends and meanders with linewidth down to 250 nm as well as inductance of vias between superconducting layers of the 8-metal layer process developed at MIT Lincoln Laboratory for fabricating VLSI superconductor circuits. We have obtained very tight inductance parameter spreads on 200-mm wafers. We have found a good agreement between the results of 3-D inductance extractors based on FastHenry [16], [22],[23] or similar numeric methods [17],[18],[30] and the experimental data, but much poorer agreement at small linewidths for 2.5-D simulators, e.g., Sonnet [31]. The data obtained can be used for SFQ circuit design into our 4M, 8M, and 10M fabrication processes. The full set of the data will also be made available as an Appendix to the Design Rules for these processes and provided to users upon request through MIT LL.

## ACKNOWLEDGMENT

We greatly appreciate discussions with Dr. William D. Oliver, Dr. Marc Manheimer, Dr. D. Scott Holmes and Dr. Anna Herr as well as their interest and support of this work.
We are also grateful to Prof. C.J. Fourie for numerous discussions of InductEx and to Prof. M. Khapaev for access to his 3-D inductor simulator.